\documentclass[11pt]{article}

\usepackage[preprint]{acl}
\usepackage[ruled,linesnumbered,vlined,algo2e]{algorithm2e}
\usepackage{algorithm}
\usepackage{placeins}

\usepackage{amssymb}

\usepackage{mathtools}

\usepackage{booktabs}
\usepackage{multirow}

\usepackage{times}
\usepackage{latexsym}

\usepackage[T1]{fontenc}

\usepackage[utf8]{inputenc}

\usepackage{microtype}
\usepackage[most]{tcolorbox}

\usepackage{inconsolata}
\usepackage{array}
\usepackage{booktabs}
\usepackage{makecell}
\usepackage{pifont}
\usepackage{tabularx}
\usepackage{xcolor}
\usepackage{colortbl}

\definecolor{tablegreen}{HTML}{2E7D32}
\definecolor{tablered}{HTML}{C2413B}
\definecolor{toolrethighlight}{HTML}{E6EEF8}

\newcommand{\cmark}{\textcolor{tablegreen}{\ding{51}}}
\newcommand{\xmark}{\textcolor{tablered}{\ding{55}}}

\usepackage[most]{tcolorbox}
\usepackage[ruled,linesnumbered,vlined,algo2e]{algorithm2e}
\usepackage{algorithm}

\usepackage{amssymb}

\usepackage{mathtools}

\usepackage{booktabs}

\usepackage{times}
\usepackage{latexsym}

\usepackage[T1]{fontenc}

\usepackage[utf8]{inputenc}

\usepackage{microtype}
\usepackage[most]{tcolorbox}

\usepackage{inconsolata}
\usepackage{cuted}
\usepackage{caption}
\usepackage{subcaption}
\usepackage{graphicx}

\usepackage[most]{tcolorbox}
\tcbuselibrary{listings,breakable}

\usepackage[most]{tcolorbox}
\usepackage{enumitem} 
\usepackage[most]{tcolorbox}
\usepackage{enumitem}

\usepackage[most]{tcolorbox}
\usepackage{enumitem}

\usepackage[most]{tcolorbox}
\usepackage{enumitem}
\lstdefinestyle{promptlisting}{
    basicstyle=\fontsize{7}{8}\selectfont\ttfamily,
    breaklines=true,
    columns=fullflexible,
    keepspaces=true,
    showstringspaces=false,
    aboveskip=0pt,
    belowskip=0pt
}

\newtcblisting{promptbox}{
    enhanced,
    breakable,
    listing only,
    listing options={style=promptlisting},
    width=\columnwidth,
    sharp corners,
    colback=white,
    colframe=black!75,
    boxrule=0.35pt,
    borderline east={0.9pt}{-1.1pt}{blue!22},
    top=1.2mm,
    bottom=1.2mm,
    left=1.5mm,
    right=1.5mm,
    before skip=4pt,
    after skip=7pt
}

\title{LIBRA: Programmatic Induction of Executable Tool Libraries}

\author{
    Zekai Li\thanks{~~Equal contribution.} \, $^1$ \quad 
    Zhenke Liu\footnotemark[1] \, $^1$ \quad 
    Yonghao Li$^1$ \quad 
    Junhan Liu$^1$ \quad 
    Xulong Xiao$^1$ \quad 
    Jiapeng Zhang$^2$ \\
    $^1$Department of Computer Science, Brown University, Providence, RI 02912, USA \\
    $^2$Department of Computer Science, University of Southern California, Los Angeles, CA 0089, USA \\
    \texttt{\{zekai\_li, zhenke\_liu, yonghao\_li, junhan\_liu, xulong\_xiao\}@brown.edu}, \quad 
    \texttt{jiapeng\_zhang@usc.edu}
}
\begin{document}
\maketitle
\begin{abstract}
Large language model agents increasingly retrieve tools from large catalogs, yet most retrievers rank each tool independently by request similarity.
For multi tool tasks, this recovers the goal matching tool while omitting prerequisite tools required by the workflow.
We introduce \textsc{LIBRA}, a framework for \emph{Library Induction by Runtime Assembly} that turns a flat tool catalog into an executable tool library.
An LLM proposes per tool contracts of candidate invocations, reusable outputs and state changes, and success conditions; \textsc{LIBRA} then tests these compositions in a sandbox and retains the relations supported at runtime.
Constructed before user queries, the library requires no dependency annotations, tool use trajectories, or task specific training, and supports both one-shot and agentic retrieval.
On AppWorld, BFCL, and SciToolEval, \textsc{LIBRA} improves one-shot Recall and NDCG over the same dense retriever by $7.9$ and $3.7$ percentage points on average.
Within ToolReAGt, it yields corresponding gains of $4.5$ and $2.9$ points after ten retrieval iterations.
\end{abstract}
\section{Introduction}
\label{sec:intro}

Large language models (LLMs) increasingly act as agents that accomplish tasks by calling external tools and APIs~\citep{toolgen,gtool}.
As tool catalogs grow, exposing every available tool to the LLM becomes costly and unreliable: the catalog may exceed the context window, while irrelevant tools increase inference cost and degrade tool selection.
Practical systems therefore retrieve a small candidate set for each user request~\citep{toolrerank,toolreagt}.

Most tool retrievers, however, represent and rank tools independently.
This abstraction is insufficient for tasks that require multiple tools to execute in sequence.
A user typically states the desired outcome rather than the complete operational procedure, leaving prerequisites such as authentication, identifier lookup, state initialization, and format conversion absent from the request.
Consequently, a retriever may identify the tool that directly matches the stated goal while omitting upstream tools whose outputs or state are required for execution.
Semantic relevance alone therefore does not guarantee a retrieval set that contains the prerequisite tools required by a valid workflow.

Existing dependency aware approaches obtain tool relations from supplied hierarchies, metadata based inference, trained dependency models, or historical tool use trajectories~\citep{toolrerank,toolnet,tgr,dtdr,gtool}.
Although these approaches can improve retrieval, they do not fully address a new executable catalog for which dependency annotations, tool-use demonstrations, and usage trajectories are unavailable.
As summarized in Table~\ref{tab:structure-comparison}, methods that learn tool structure either depend on existing tool use trajectories or training, or accept metadata inferred relations without testing whether the tools actually compose.
This distinction matters because a relation can appear plausible from tool metadata yet fail in the target runtime.
We therefore investigate the following question: How can we construct reusable knowledge of tool composition before user queries or usage histories become available?

\begin{table*}[t]
\centering
\small
\setlength{\tabcolsep}{14pt}
\renewcommand{\arraystretch}{1.15}
\caption{
Comparison of methods for learning tool dependencies.
Prior methods either depend on existing tool use trajectories or training, or retain metadata inferred relations without runtime validation.
\textsc{LIBRA} performs trajectory free, training free dependency induction and validates candidate tool compositions through execution in the target environment.
}
\label{tab:structure-comparison}
\begin{tabular}{
    @{}
    l
    >{\centering\arraybackslash}p{0.25\textwidth}
    >{\centering\arraybackslash}p{0.14\textwidth}
    >{\centering\arraybackslash}p{0.20\textwidth}
    @{}
}
\toprule
Method
& Trajectory free
& Training free
& Execution validation \\
\midrule

ToolNet~\citep{toolnet}
& \xmark
& \cmark
& \xmark \\

TGR~\citep{tgr}
& \cmark
& \xmark
& \xmark \\

GTool~\citep{gtool}
& \xmark
& \xmark
& \xmark \\

Ego-Trip~\citep{egotrip}
& \cmark
& \cmark
& \xmark \\

\midrule

\textbf{\textsc{LIBRA}}
& \cmark
& \cmark
& \cmark \\

\bottomrule
\end{tabular}
\end{table*}

We propose \textsc{LIBRA}, a framework for \emph{Library Induction by Runtime Assembly}.\footnote{Code will be released upon acceptance.}
We distinguish a tool catalog, which contains callable interfaces
and public metadata, from an executable tool library, which augments
the catalog with per tool executable contracts and runtime supported
composition relations.
Each contract specifies candidate invocations, reusable outputs and state changes, and observable success conditions.
\textsc{LIBRA} executes type compatible compositions in a sandbox and can retain input provenance relations, compatibility relations, and state dependencies supported by constructing counterfactual tests.
Each test checks not merely that a composition succeeds, but that it fails without the candidate dependency; this yields relations that are more precise than metadata inferred structure (Appendix~\ref{app:ablation}). 
The resulting query independent library augments tool retrieval with reusable composition knowledge, but does not generate or execute task specific plans.
We therefore evaluate its ability to recover tools required under each benchmark's evaluation protocol.

The executable tool library is constructed ahead of user queries and requires neither existing tool use trajectories nor task specific training.
At retrieval time, a dense base retriever identifies semantically
relevant seed tools, while the library promotes their prerequisite
tools, which need not be mentioned in the request.
Because the library is query independent, the same artifact can augment both one-shot retrievers and agentic retrievers that search iteratively.
\textsc{LIBRA} thus learns how tools compose once and exposes that knowledge to subsequent retrieval workflows.

We evaluate \textsc{LIBRA} on AppWorld, BFCL, and SciToolEval against lexical, dense, representation augmented, graph based, and agentic retrieval baselines.
Relative to the same dense retriever, \textsc{LIBRA} improves one-shot Recall and NDCG by $7.9$ and $3.7$ percentage points on average across the three benchmarks.
When integrated into ToolReAGt, \textsc{LIBRA} improves Recall and NDCG by $4.5$ and $2.9$ percentage points, averaged across three benchmarks, three agent LLMs, and three random seeds. 
These results demonstrate that an execution supported tool library provides reusable retrieval guidance beyond semantic similarity alone.

Our contributions are as follows:
\begin{itemize}
    \item We introduce \textsc{LIBRA}, which uses LLM proposed executable contracts and a counterfactual sandbox execution test, retaining a relation only when the provider value succeeds while an independently constructed control fails, to construct a reusable tool library of runtime supported composition relations without dependency annotations, existing tool use trajectories, or task specific training.

    \item We demonstrate that the induced tool library improves both one-shot and agentic retrieval across three tool use benchmarks and three agent LLMs.

    \item We present case studies of prerequisite tools that \textsc{LIBRA} recovers despite their low semantic similarity to the request, together with a breakdown of the remaining causes of failed dependency recovery.
\end{itemize}

\section{Related Work}

\begin{figure*}[t]
    \centering
    \includegraphics[width=\linewidth]{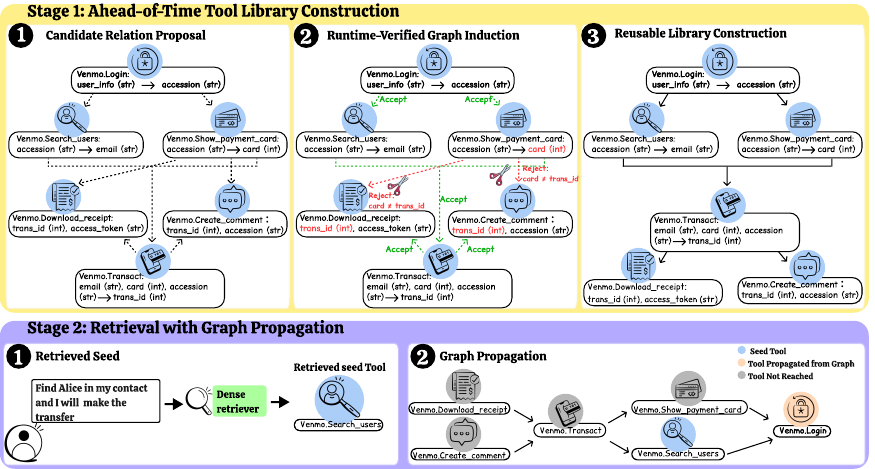}
    \caption{
    Overview of \textsc{LIBRA}.
    \textbf{Stage 1:} Before user queries arrive, an LLM proposes an executable contract of each tool.
    \textsc{LIBRA} tests each candidate composition in the target runtime.
    Successful compositions are retained to construct the executable tool library, while unsuccessful ones are removed.
    For example, although \texttt{card} and \texttt{trans\_id} share the same integer type, a payment card identifier cannot be used as a transaction identifier.
    \textbf{Stage 2:} Given the request ``Find Alice in my contact and I will make the transfer,'' dense retrieval returns \texttt{Venmo.Search\_users}.
    \textsc{LIBRA} performs one-hop propagation and recovers its prerequisite tool \texttt{Venmo.Login}.
    }
    \label{fig:libra-overview}
\end{figure*}

\paragraph{Tool representations and structural retrieval.}
The dominant approach to large tool catalogs represents and ranks tools independently.
Sparse retrievers instantiate this approach with lexical matching~\citep{bm25,tfidf}, while dense retrievers use embedding similarity.
Representation augmentation methods enrich either the query or each tool's description: PLUTO decomposes requests and edits tool descriptions~\citep{pluto};
ReInvoke generates synthetic indexing queries and rewrites invocation intent at inference time~\citep{reinvoke};
and ToolREX constructs structured tool profiles~\citep{toolrex}.
Parametric methods instead internalize tool identity: ToolGen assigns a learned token to each tool~\citep{toolgen}, whereas ToolWeaver learns hierarchical relationships among tools from semantics and existing tool use trajectories~\citep{toolweaver}.

Structure aware approaches differ along two axes: whether they support retrieval without target domain usage traces and whether structure induction requires training.
ToolRerank relies on ToolBench's existing API hierarchy and trains a retriever, reranker, and query type classifier~\citep{toolrerank}.
TGR constructs a new catalog graph from tool metadata using a dependency discriminator trained on synthetic and manually labeled tool pairs~\citep{tgr}.
GTool derives edges from historical tool use trajectories and trains a GNN for missing-edge prediction and tool planning~\citep{gtool}.
Ego-Trip supports deployment without target domain usage traces or parameter training, but infers relations solely from tool metadata through LLM prompting~\citep{egotrip}.

\textsc{LIBRA} requires neither prior trajectories nor task specific
training, as it induces composition relations from successful executions in the target runtime, retaining a relation only when a counterfactual test confirms that the composition succeeds with the candidate dependency and fails without it.
The resulting reusable executable library improves both one-shot and
agentic retrieval.
Table~\ref{tab:structure-comparison} summarizes these differences.

\paragraph{Agentic retrieval.}
A second family integrates tool retrieval into agentic workflows.
Iterative Feedback repeatedly asks the tool using LLM to refine the retriever's view of the query and candidates~\citep{iterfeedback}.
ToolReAGt uses ReAct prompting to decompose requests and retrieve relevant tools over multiple rounds~\citep{toolreagt,react}.
ToolQP trains a query planner on synthetic trajectories using supervised learning and reinforcement learning with verifiable rewards to decompose requests and perform retrieval~\citep{toolqp}.
DTDR estimates next tool structure from existing trajectories and demonstrations to condition retrieval on both the request and tool call history~\citep{dtdr}.
These methods can recover tools omitted by one-shot retrieval, but perform adaptive reasoning separately for each request;
ToolQP and DTDR additionally depend on training or prior tool use histories.
\textsc{LIBRA} instead performs training free, ahead-of-time structural augmentation, converting a flat catalog into a reusable executable tool library containing compositional relations before queries arrive.
We further show that the resulting library improves agentic retrieval.

\section{Method}
\label{sec:method}
Given a tool catalog \(\mathcal{T}=\{t_1,\ldots,t_n\}\) and its runtime, \textsc{LIBRA} constructs an executable tool library
\begin{equation}
    \mathcal{L}=(\mathcal{T},\mathcal{C},\mathcal{E}),
\end{equation}
where $\mathcal{C} = \{c_i\}_{i=1}^n$ contains an executable contract
for each tool, and $\mathcal{E}$ contains the relations retained by the composition tests below.
Here, \emph{runtime-validated} means that a relation passes its specified test under the tested contract and sandbox state; it is not a universal guarantee over all possible inputs or states.
For an edge \(t_p\!\rightarrow\!t_c\), \(t_p\) is the \emph{provider}, whose execution produces a value or establishes a state, and \(t_c\) is the \emph{consumer}, whose execution uses that value or state.
At retrieval time, only original tool metadata will be embedded to provide representations, while the relations admitted to the benchmark specific retrieval graph support propagation. 
Algorithms~\ref{alg:library-construction} and~\ref{alg:graph-propagation} in Appendix~\ref{sec:algorithms} provide pseudocode for library construction and one-hop graph propagation, respectively.

\subsection{Ahead-of-Time Library Construction}
\label{subsec:construction}

\paragraph{Executable Contracts Generation.}
\textsc{LIBRA} processes each tool independently using only its public description, signature, and input and output schema.
An LLM produces a contract
\begin{equation}
    c_i=(\Pi_i,\mathcal{X}_i,S_i^{+},S_i^{-}),
\end{equation}
where \(\Pi_i\) is a set of candidate calls, \(\mathcal{X}_i\) specifies reusable values from the call arguments, returned
response, or runtime state changes, and \(S_i^{+}\) and \(S_i^{-}\) describe observable success and failure outcomes.
Each candidate call \(\pi\in\Pi_i\) specifies how every input parameter is filled and whether the call is expected to succeed or fail. 
A parameter may receive a proposed literal, obtain a value from an earlier tool call, or be omitted when it is optional.

\textsc{LIBRA} executes the candidate calls in each contract in isolated sandbox states.
If an observed result matches neither outcome specification, the LLM receives the execution feedback and revises the contract.
For each tool, \textsc{LIBRA} makes one initial contract-generation call and at most two feedback-based revision calls.
Thus, for a catalog of \(n\) tools, the total number of LLM calls is linear in \(n\).

\paragraph{Runtime Relation Induction.}
\textsc{LIBRA} tests whether an observed value produced by a provider tool $t_p$ can be used as an input to a consumer tool $t_c$.
Each relation is admitted by a \emph{counterfactual} test rather than a single success check, which separates genuine dependencies from merely schema compatible pairs.
\textsc{LIBRA} first executes $t_p$, extracts an output declared in the contract, and supplies that value to a compatible input of $t_c$.
If the invocation succeeds, \textsc{LIBRA} invokes $t_c$ again from the same saved sandbox state, replacing that input with an independently constructed literal or omitting it when the input is optional.
If the call with the observed provider value succeeds while the literal-or-omission control fails, \textsc{LIBRA} retains an \emph{input-provenance relation} $t_p \rightarrow t_c$: the upstream value is necessary, relative to the tested control, for that consumer invocation to succeed under the fixed sandbox state.
State-mediated prerequisites are assessed separately by the before--after state control.
If both invocations succeed, the connection is retained as a compatibility relation only when the provider output and consumer input are also semantically aligned.

Some tools communicate through environment state rather than explicit values.
\textsc{LIBRA} tests these cases by invoking the consumer \(t_c\) with identical arguments before and after executing a candidate provider \(t_p\).
A state dependency \(t_p\!\rightarrow\!t_c\) is retained only when the
consumer fails before \(t_p\) is executed and succeeds afterward.
Observed values and states produced by successful calls are reused in
later invocations, allowing \textsc{LIBRA} to discover multi-hop
compositions.

\subsection{Retrieval with Graph Propagation}
\label{subsec:library-retrieval}

Given a query \(q\), the base retriever assigns each tool a normalized dense relevance score
\begin{equation}
    s_q(t)=\frac{1+\cos(h(q),h(t))}{2},
\end{equation}
where \(h\) is the dense encoder applied to the original tool metadata and the executable contracts are not embedded as retrieval documents. The retriever selects the top \(K_s\) tools as consumer seeds. Retrieval propagates over every relation admitted to the
benchmark-specific retrieval graph, including compatibility relations
when they are retained.

\begin{table*}[t]
\centering
\small
\renewcommand{\arraystretch}{1.12}
\caption{
One-shot retrieval results without query time reranking.
Dataset@\(\!K\) denotes the evaluation cutoff.
Values are percentages.
The best result in each column is highlighted, and the second best is underlined.
}
\label{tab:one-shot-results}
\begin{tabular*}{\textwidth}{
@{\extracolsep{\fill}}
l
ccc
ccc
ccc
@{}
}
\toprule
& \multicolumn{3}{c}{AppWorld@15}
& \multicolumn{3}{c}{SciToolEval@10}
& \multicolumn{3}{c}{BFCL@15} \\
\cmidrule(lr){2-4}
\cmidrule(lr){5-7}
\cmidrule(lr){8-10}
Method
& P & R & NDCG
& P & R & NDCG
& P & R & NDCG \\
\midrule
BM25~\citep{bm25}
& 10.9 & 20.0 & 18.8
& 12.0 & 86.4 & 54.1
& 26.2 & 75.5 & 69.5 \\

QTS
& 15.5 & 28.7 & 32.8
& 11.3 & 84.1 & \underline{83.9}
& 28.6 & 81.8 & 76.2 \\

ToolREX~\citep{toolrex}
& 14.5 & 26.8 & 30.7
& 11.1 & 83.1 & 83.8
& 27.1 & 77.7 & 71.9 \\

TGR-Adapt~\cite{tgr}
& \underline{19.7}
& \underline{36.3}
& \cellcolor{toolrethighlight}\textbf{39.9}
& 10.2 & 76.2 & 75.1
& 28.5 & 81.4 & 73.6 \\

Ego-Trip~\citep{egotrip}
& 15.8 & 29.2 & 33.2
& 11.3 & 84.1 & \underline{83.9}
& \underline{30.4}
& \underline{85.7}
& \cellcolor{toolrethighlight}\textbf{78.4} \\

SciToolKG~\citep{scitoolagent}
& -- & -- & --
& \underline{12.6}
& \underline{89.7}
& 82.5
& -- & -- & -- \\

\addlinespace[1pt]
\textbf{\textsc{LIBRA}}
& \cellcolor{toolrethighlight}\textbf{21.8}
& \cellcolor{toolrethighlight}\textbf{40.7}
& \underline{39.1}
& \cellcolor{toolrethighlight}\textbf{12.7}
& \cellcolor{toolrethighlight}\textbf{91.1}
& \cellcolor{toolrethighlight}\textbf{86.6}
& \cellcolor{toolrethighlight}\textbf{30.6}
& \cellcolor{toolrethighlight}\textbf{86.4}
& \cellcolor{toolrethighlight}\textbf{78.4} \\
\bottomrule
\end{tabular*}
\end{table*}

Starting from each seeded consumer, \textsc{LIBRA} follows incoming
graph edges for one hop to its providers.
For each input of a seeded consumer, \textsc{LIBRA} retains at most two providers: the provider with the highest output-to-input field similarity and the provider with the highest dense relevance \(s_q\).
If the same provider satisfies both criteria, it is retained only once. 
This pruning recovers plausible prerequisites without adding every provider that passes the schema check to the candidate set.

Let \(C_q(u)\) denote the set of distinct seeded consumers connected to provider \(u\).
Each consumer propagates its relevance score \(s_q(t)\) to \(u\).
\textsc{LIBRA} averages the provider's original relevance with these consumer scores:
\begin{equation}
    g_q(u)=
    \frac{s_q(u)+\sum_{t\in C_q(u)}s_q(t)}
    {1+|C_q(u)|}.
\end{equation}
\textsc{LIBRA} then sets
\begin{equation}
    s'_q(u)=\max\bigl(s_q(u),g_q(u)\bigr).
\end{equation}

A consumer contributes at most once to a provider, even when several of its inputs select the same provider.
Averaging prevents a tool from receiving a higher score solely because it has more graph neighbors.
Tools outside the retrieval graph retain their original dense scores.
\textsc{LIBRA} selects the top \(K_c\) tools under \(s'_q\) as the candidate set.

\section{Evaluation}
\label{sec:experiments}
We evaluate \textsc{LIBRA} as a tool-retrieval method.
Specifically, we ask \textbf{whether the induced library improves the recovery and ranking of tools in one-shot retrieval} and \textbf{whether the same library remains useful for agentic retrieval}.
Retrieval metrics measure candidate-set quality rather than downstream
task completion.

\subsection{Experimental Setup}
\label{subsec:setup}

\paragraph{Benchmarks.}
We evaluate on three benchmarks spanning different catalog sizes, domains, and composition patterns:
AppWorld~\citep{appworld} covers everyday app workflows;
SciToolEval~\citep{scitoolagent} targets scientific computing pipelines;
and the multi turn base split of BFCL~\citep{bfcl} contains function calling tasks over filesystem and utility APIs.
We convert each benchmark into a common retrieval format comprising a query, a benchmark wide tool catalog, and a gold set of required tools.
For each benchmark, \textsc{LIBRA} constructs one library using only public tool metadata and the executable runtime; it does not access task instructions, gold tool sets, or split labels during construction.
Required tool annotations are used only for post retrieval evaluation, and no task level annotations from these benchmarks are used to select retrieval settings.
Further details are provided in Appendix~\ref{sec:data}.

\paragraph{Baselines.}
We compare against following sparse and dense retrieval methods.
\textbf{BM25}~\citep{bm25} and \textbf{Query Tool Similarity} (QTS) represent lexical and dense retrieval without structural augmentation.
\textbf{ToolREX}~\citep{toolrex} augments each tool with an LLM generated profile before embedding.
\textbf{TGR-Adapt}~\citep{tgr} and \textbf{Ego-Trip}~\citep{egotrip} both construct tool graphs from metadata via LLM based dependency classification and relation extraction, respectively.
We additionally include SciToolKG, a knowledge graph retriever, as a baseline~\citep{scitoolagent}.
We do not compare against methods that require existing tool use trajectories or task specific training as they are inapplicable to our setting.
For agentic retrieval, we integrate each method into \textbf{ToolReAGt}~\citep{toolreagt} with three agent LLMs: Gemma-4-12B, Nemotron-9B, and Qwen3.5-9B~\citep{gemma4,nemotron,qwen3.5}.
All prompts are in Appendix~\ref{sec:prompts}.

\begin{figure*}[t]
    \centering
    \includegraphics[width=0.9\textwidth]{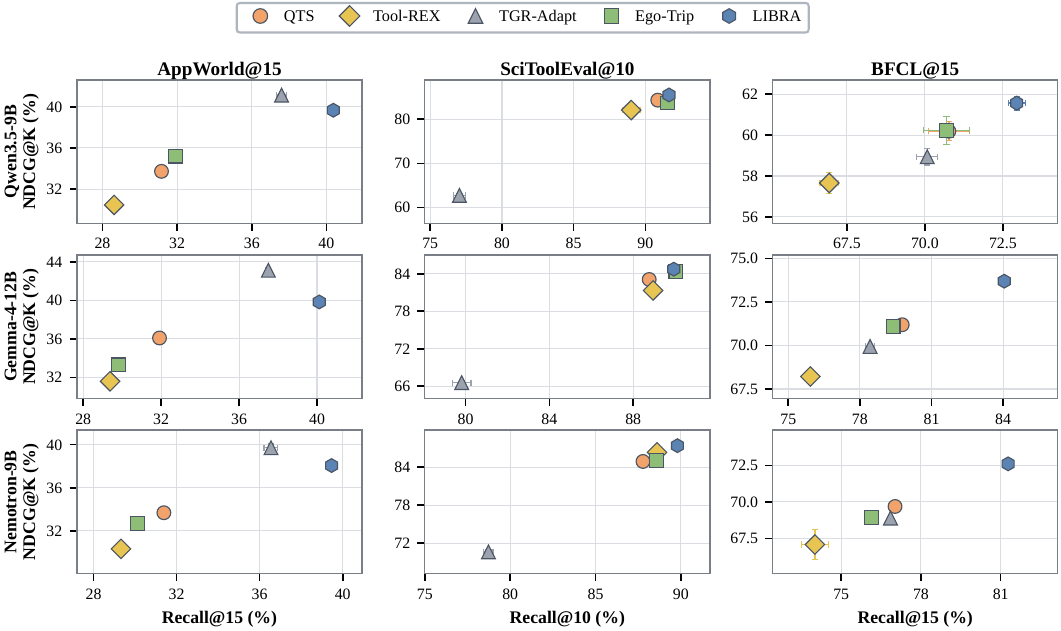}
    \caption{
    Agentic retrieval after $10$ iterations.
    Each panel reports Recall and NDCG for one agent LLM and benchmark.
    Points show the mean over three random seeds, and error bars show standard error.
    \textsc{LIBRA} attains the highest Recall in $8$ of the $9$ settings and the highest NDCG on SciToolEval and BFCL for all three agent LLMs.
    }
    \label{fig:agentic-retrieval}
\end{figure*}

\paragraph{Metrics and protocol.}
We report Precision, Recall, and NDCG at \(K\!=\!15\) for AppWorld and BFCL and \(K\!=\!10\) for SciToolEval.
All embedding based methods use Qwen-3 Embedding 0.6B~\citep{qwen3embedding_rerank}, return at most 20 candidates, and are evaluated without query time reranking.

\paragraph{Library construction.}
\textsc{LIBRA} constructs the executable tool library once per catalog using DeepSeek-v4-Flash~\citep{dsv4} for contract generation.
The library is constructed once per benchmark and reused across all
\textsc{LIBRA} queries and all agent LLMs.
We report additional statistics such as catalog size in Appendix~\ref{sec:construction-statistics}.

\subsection{One-Shot Retrieval}
\label{subsec:one-shot-results}

Table~\ref{tab:one-shot-results} compares \textsc{LIBRA} with one-shot retrieval baselines.
Relative to QTS, which uses the same embedding model without the induced retrieval graph, \textsc{LIBRA} improves Recall and NDCG by $7.9$ and $3.7$ percentage points on average across the three benchmarks.

\textbf{On SciToolEval, \textsc{LIBRA} outperforms the official knowledge graph retriever SciToolKG} by $1.4$ and $4.1$ points for Recall and NDCG.
The largest gains occur on AppWorld, whose tasks require 8.18 tools on average: \textsc{LIBRA} improves Recall by $12$ points and NDCG by $6.3$ points over QTS.

\subsection{Agentic Retrieval}
\label{subsec:agentic-results}

Compared with QTS, which uses the same dense retrieval backend without the executable tool library, \textsc{LIBRA} improves Recall and NDCG by $4.5$ and $2.9$ percentage points, averaged across three benchmarks, three agent LLMs, and three random seeds. Averaged across LLMs and seeds, the gains are $8.5$ Recall and $4.7$ NDCG points on AppWorld, $1.3$ and $1.8$ points on SciToolEval, and $3.6$ and $2.3$ points on BFCL.

\textsc{LIBRA} achieves the highest Recall for every agent LLM on AppWorld and BFCL. On SciToolEval, it achieves the highest Recall with Qwen3.5 and Nemotron and trails Ego-Trip by only $0.1$ points with Gemma-4. It also achieves the highest NDCG for every agent LLM on SciToolEval and BFCL. Although TGR-Adapt obtains higher NDCG on AppWorld, \textsc{LIBRA} retrieves a larger fraction of the required tools. These results show that the same query independent executable library remains effective as agents reformulate their searches over multiple rounds, without modifying the ReAct policy or training the agent LLM.

\section{Analysis of Dependency Recovery}
\label{sec:analysis}

\subsection{Dependencies Bridge Semantic Gaps}
\label{subsec:semantic-gap}

\begin{figure*}[t]
    \centering
    \begin{subfigure}[t]{0.32\textwidth}
        \centering
        \includegraphics[width=\linewidth]{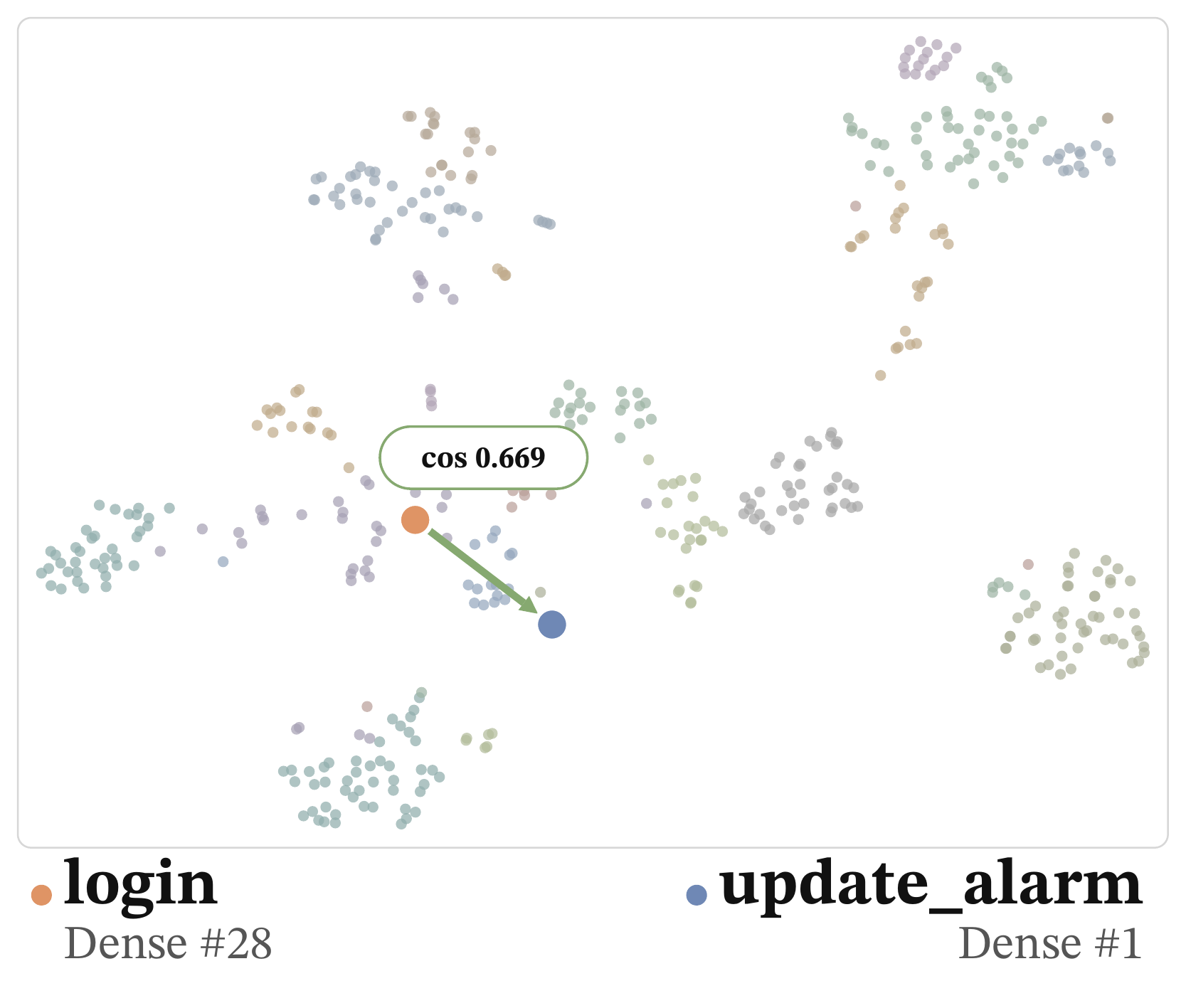}
        \caption{AppWorld}
    \end{subfigure}
    \hfill
    \begin{subfigure}[t]{0.32\textwidth}
        \centering
        \includegraphics[width=\linewidth]{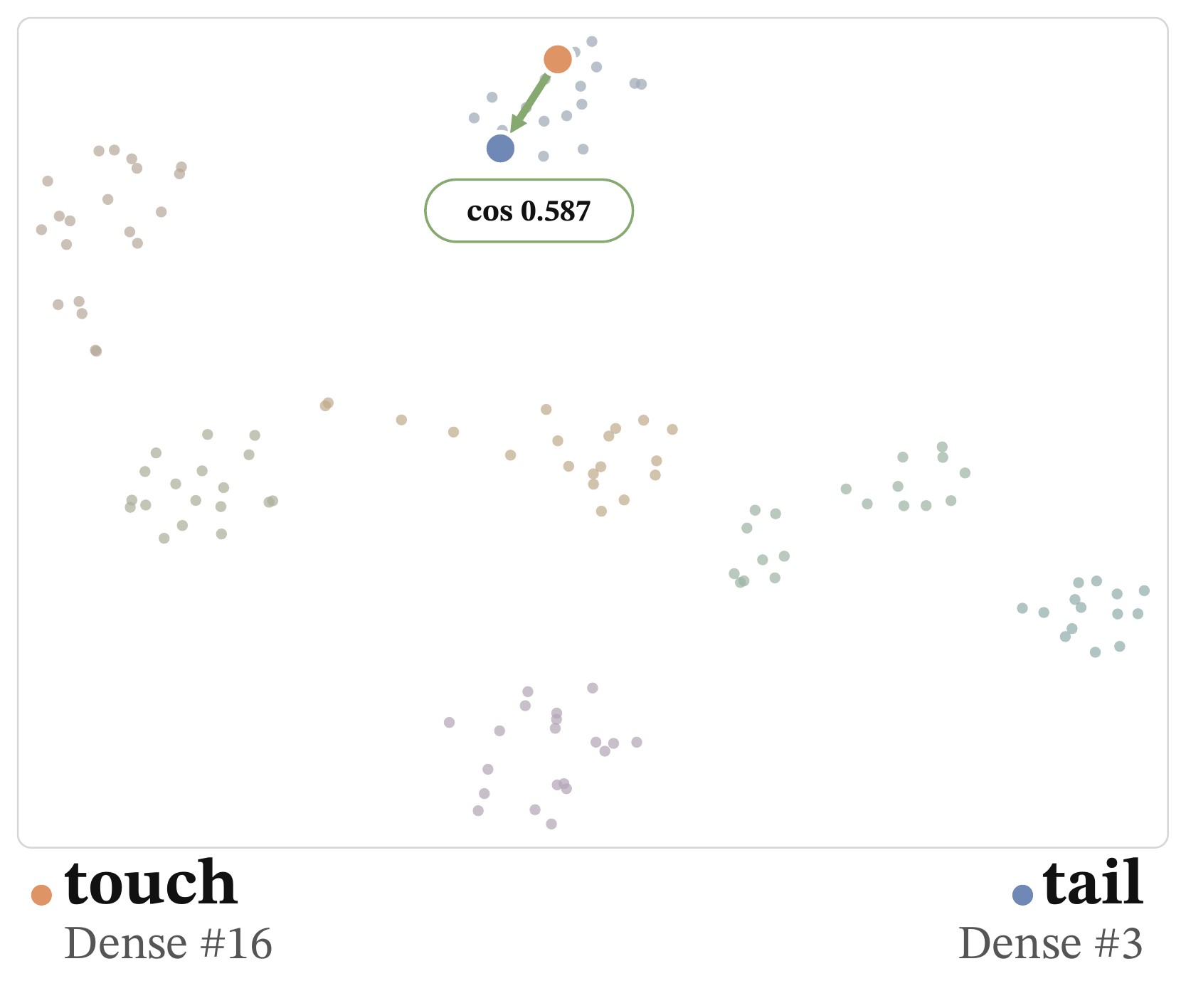}
        \caption{BFCL}
    \end{subfigure}
    \hfill
    \begin{subfigure}[t]{0.32\textwidth}
        \centering
        \includegraphics[width=\linewidth]{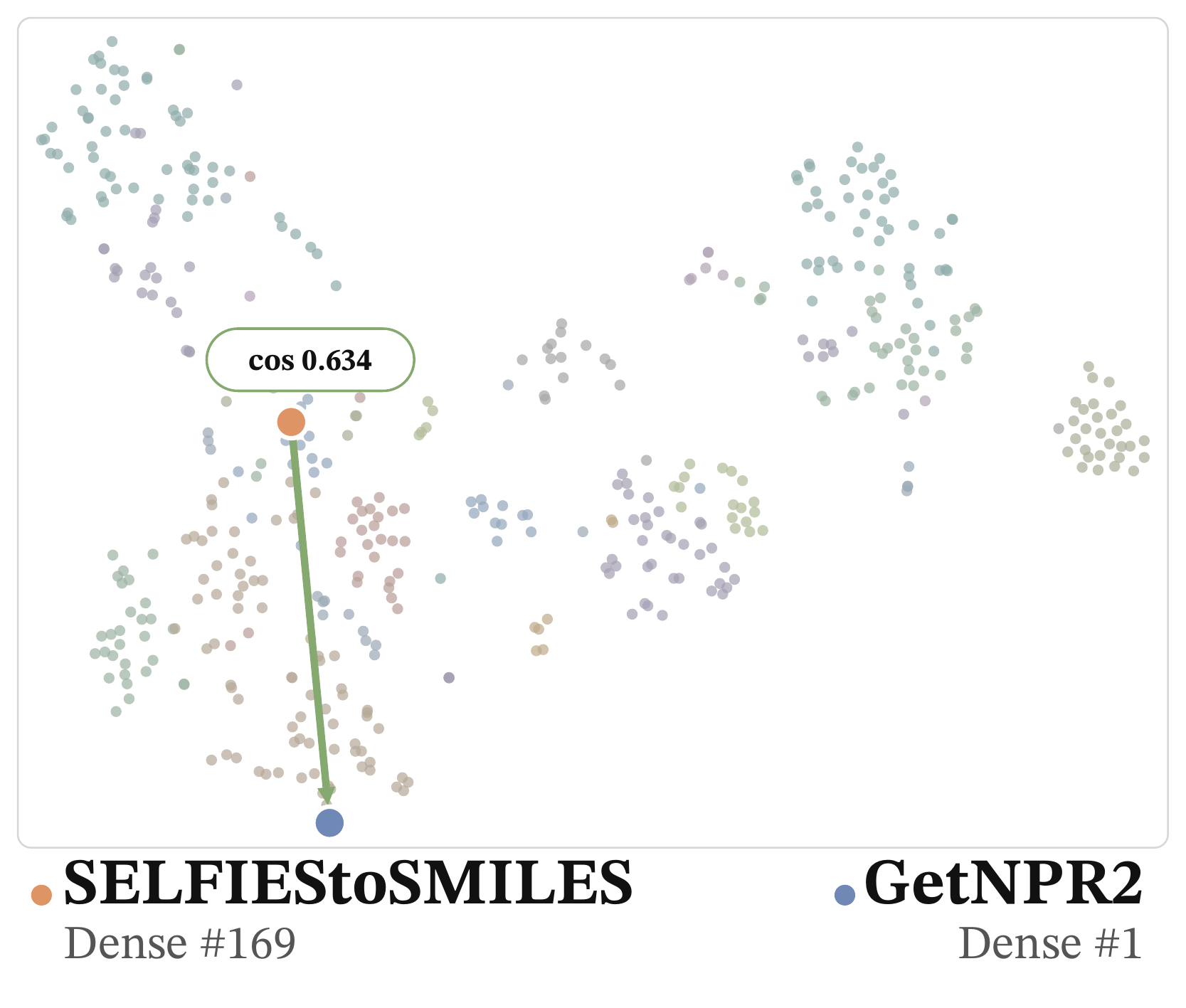}
        \caption{SciToolEval}
    \end{subfigure}
\caption{
We projected tool embeddings separately for each benchmark using t-SNE, with background colors indicating different clusters.
In each example, Dense retrieval ranks the consumer (blue) within the top \(K\) but misses its provider (orange):
(a) \texttt{phone.update\_alarm} (\#1) depends on \texttt{phone.login} (\#28);
(b) \texttt{tail} (\#3) depends on \texttt{touch} (\#16); and
(c) \texttt{GetNPR2} (\#1) depends on \texttt{SELFIEStoSMILES} (\#169).
\textsc{LIBRA} follows the retained relation and recovers the provider.
For SciToolEval and AppWorld, we observed runtime dependencies connecting tools across different clusters.
}

    \label{fig:dependency-semantic-gap}
\end{figure*}

The experimental results show that \textsc{LIBRA} retrieves more required tools.
They leave two questions open: why a dependency library helps and why recovery still fails in some cases.
Following the evaluation setup, we examine the frozen one-shot rankings and induced libraries using \(K=15\) for AppWorld and BFCL and \(K=10\) for SciToolEval.

Figure~\ref{fig:dependency-semantic-gap} shows three representative dependencies recovered by \textsc{LIBRA} despite low semantic similarity between each provider and the request.
The AppWorld request asks to move an alarm, placing \texttt{phone.update\_alarm} at Dense rank 1.
The authentication prerequisite is implicit, leaving \texttt{phone.login} at rank 28.
In BFCL, a request to read the tail of a newly created CSV directly matches \texttt{GorillaFileSystem.tail}.
Its state creating prerequisite \texttt{GorillaFileSystem.touch} is ranked 16.
The rank gap is largest on SciToolEval: a request for the NPR2 descriptor retrieves \texttt{GetNPR2} at rank 1, whereas the required format converter \texttt{SELFIEStoSMILES} is ranked $169$.
In all three cases, \textsc{LIBRA} moves the provider into the top \(K\).

These examples represent three forms of composition: authentication, environment state, and data format conversion.
The semantic embeddings of the provider and consumer tools need not be similar because the tools implement distinct operations.
Dense retrieval therefore behaves as intended when it finds the consumer that directly matches the request; the missing signal is the dependency that specifies which prerequisite must be executed first.
The induced library supplies this dependency signal without forcing prerequisite tools to resemble the query or their consumers in embedding space.

\subsection{Dependency Recovery Breakdown}
\label{subsec:error-audit}

We analyze gold tool occurrences ranked outside the top \(K\) by query tool similarity.
We assign each occurrence to one of five mutually exclusive outcomes.
If \textsc{LIBRA} moves the provider into the top \(K\), the occurrence is marked as \emph{recovered}.
Otherwise, the occurrence is assigned to one of four failure causes: the provider has no outgoing edge in the retrieval graph; none of its connected consumers is retrieved as a seed; it is not selected among competing providers; or its propagated score remains below the cutoff.


\begin{table}[hbtp!]
\centering
\renewcommand{\arraystretch}{1.12}
\caption{
Recovery outcomes for query specific gold tool occurrences outside the top \(K\) under Dense retrieval.
Each occurrence is assigned to exactly one outcome; percentages sum to 100 within each benchmark.
}
\label{tab:failure-attribution}
\resizebox{\columnwidth}{!}{%
\begin{tabular}{lrrrrrr}
\toprule
Benchmark
& \makecell{Dense misses\\($n$)}
& \makecell{No graph\\edge (\%)}
& \makecell{Consumer not\\seeded (\%)}
& \makecell{Provider not\\selected (\%)}
& \makecell{Below\\cutoff (\%)}
& \makecell{Recovered\\(\%)} \\
\midrule
AppWorld    & 860 & 65.1 & 13.0 & 3.4  & 0.2  & 18.3 \\
BFCL        & 224 & 2.7  & 30.4 & 12.1 & 21.4 & 33.5 \\
SciToolEval & 85  & 38.8 & 0.0  & 2.4  & 0.0  & 58.8 \\
\bottomrule
\end{tabular}%
}
\end{table}

Table~\ref{tab:failure-attribution} reveals different causes of failed recovery across benchmarks.
SciToolEval has the highest recovery rate: \textsc{LIBRA} recovers 58.8\% of gold tool occurrences ranked outside the top \(K\) by query tool similarity, and almost every remaining miss has no edge in the retrieval graph.
On BFCL, most failures occur later in the recovery pipeline.
Only 2.7\% of the Dense misses have no outgoing graph edge; for 30.4\%, no connected consumer is retrieved as a seed; 12.1\% are not selected among competing providers; and 21.4\% are promoted but remain below the cutoff.
Seed retrieval and ranking are therefore the main causes of failed dependency recovery on BFCL.

\subsection{Failure Case Studies}
\label{subsec:failure-modes}

We examine one representative example for each failure category in Table~\ref{tab:failure-attribution}.
These examples show how graph coverage, seed retrieval, provider selection, and score propagation can each limit dependency recovery.

\paragraph{No graph edge.}
The corresponding relation may be absent because ahead-of-time
construction cannot execute a prerequisite that requires a
query specific input.
For example, \texttt{phone.search\_contacts} must retrieve a recipient's \texttt{phone\_number} for \texttt{phone.send\_text\_message}.
At execution time, the environment builds the active task database from shared base data, applies task specific changes, and sets the requesting user.
The query then specifies which contact to retrieve.
During \textsc{LIBRA}'s ahead-of-time construction, however, contracts are generated from public tool metadata without access to the task instruction.
As a result, executing \texttt{phone.search\_contacts} produces no \texttt{phone\_number} for the downstream consumer.
Without this output, \textsc{LIBRA} cannot retain the corresponding
input provenance relation between \texttt{phone.search\_contacts} and \texttt{phone.send\_text\_message}.

\paragraph{Consumer not seeded.}
A retained relation cannot contribute to retrieval when its consumer
is not retrieved as a seed.
For example, the induced graph contains an edge from the gold prerequisite \texttt{login} in the filesystem API to the gold consumer \texttt{create\_file}.
However, the provider \texttt{login} and consumer \texttt{create\_file} have Dense ranks 99 and 387, respectively.
Because \texttt{create\_file} is not retrieved as a seed, the graph edge is never activated and cannot improve retrieval.

\paragraph{Provider not selected.}
A valid provider is excluded when competing providers score higher under both per input selection criteria.
For example, a query asks to follow every artist appearing in the user's liked songs and albums.
The gold plan uses \texttt{show\_album} to obtain the \texttt{artist\_id} required by \texttt{follow\_artist}, and the induced graph contains this edge.
However, \texttt{show\_following\_artists} also connects to the same input and has higher output input field similarity (0.928 versus 0.783) and higher query relevance (Dense rank 2 versus rank 31).
Consequently, \texttt{show\_album} satisfies neither per input selection criterion, is not selected, and receives no propagated score.

\paragraph{Below cutoff.}
A selected provider can remain unrecovered when score propagation does not lift it above the retrieval cutoff.
For example, the vehicle API's \texttt{startEngine} tool is a Dense seed at rank 10, and the library contains a state dependency from \texttt{lockDoors} to \texttt{startEngine}.
Propagation promotes \texttt{lockDoors} from rank 55 to rank 35, which remains below the cutoff.

\section{Conclusion}
We introduced \textsc{LIBRA} to transform a flat executable tool catalog into a reusable library of runtime-supported composition relations. From LLM-proposed contracts, \textsc{LIBRA} admits a relation only when a counterfactual sandbox test confirms it, and propagates the validated relations to recover workflow prerequisites omitted from user requests. Construction uses CPU resources but requires no GPU for model training, with a number of LLM calls linear in the catalog size. Once constructed, the library can be deployed across tasks: it improves one-shot retrieval over dense baselines and remains effective within ToolReAGt across benchmarks and agent LLMs, showing that runtime-supported composition complements semantic similarity. By separating catalog-level composition knowledge from query-time reasoning, \textsc{LIBRA} enables reuse without query-specific reconstruction.

\section{Limitations}
\textsc{LIBRA} requires a benchmark whose tools are actually executable in an isolable, resettable runtime; many synthetic tool use benchmarks provide only tool signatures or mocked responses, to which \textsc{LIBRA} does not apply. Construction also incurs an upfront execution cost, though it amortizes across queries.

The counterfactual test governs the \emph{precision} of retained relations, not the \emph{coverage} of relations proposed in the first place. Because ahead-of-time construction cannot observe the query specific environment, some prerequisites cannot be exercised during construction, leaving their relations absent before any test applies. This gap is benchmark dependent: it dominates on AppWorld but is negligible on BFCL (Table~\ref{tab:failure-attribution}).. A retained relation also cannot help if its consumer is not retrieved as a seed.
\bibliography{custom}
\appendix
\clearpage
\section{Prompts}
\label{sec:prompts}
\raggedbottom
We report the LLM prompts used in all experiments, including \textsc{LIBRA} contract generation, Tool-REX profile construction, TGR and Ego-Trip graph construction, and ToolReAGt agentic retrieval.

\textsc{LIBRA} uses the following prompt for initial contract generation and runtime feedback revisions.
For each tool, one initial generation call and at most two revisions share a three-call budget.
Braced terms denote values serialized directly from the tool metadata, runtime feedback, or the contract schema.

\begin{promptbox}
# System
Write an executable call contract for one tool from its public metadata and supplied same-tool feedback only.

# User
{
  "contract_version": "tool_call_contract_v2",
  "revision": {revision},
  "feedback_policy": {feedback_policy},
  "public_tool_metadata": {public_tool_metadata},
  "invocation_record_schema": {invocation_record_schema},
  "required_output_schema": {tool_call_contract_schema},
  "rules": {contract_rules},
  "previous_contract": {previous_contract},
  "instruction": "Return one complete replacement contract.",
  "same_tool_invocation_feedback": {same_tool_feedback},
  "feedback_note": {feedback_note}
}
\end{promptbox}
\captionof{figure}{\textsc{LIBRA} executable contract generation prompt. Braced fields are populated from the tool metadata, contract schema, or same-tool runtime feedback.}
\label{prompt:libra-contract}

The last four fields are included only during revision or when same-tool execution feedback is available.
The complete shared rule list used in \texttt{contract\_rules} is shown below.

\begin{promptbox}
1. Describe exactly one supplied tool and return only the complete JSON contract.
2. Represent every required public parameter in every call template.
3. Use literal when a caller can construct the value from public information alone.
4. Use observation when successful use needs an exact value learned at runtime.
5. Dependency-required is independent of whether the public signature marks a parameter required: use observation exactly when this call must refer to a particular pre-existing value that a caller cannot choose or derive from public information.
6. A schema-valid placeholder or guessed identifier, token, handle, or reference is not a constructible literal when an arbitrary same-type value would fail; never use a dummy value merely because it has the right JSON type.
7. Every literal in an expected=success template must be a plausible value that can work as written using public information alone; otherwise use observation.
8. A success template may and should contain multiple observation arguments when successful use needs multiple exact runtime values.
9. Test constructibility against success: if replacing an input with another arbitrary schema-valid value would require different pre-existing setup or make the call fail, that input is not independently constructible and must use observation.
10. Inputs that this call freely chooses for a new value---such as initial attributes, a newly chosen password, user-authored content, amounts, quantities, dates, flags, and search or filter criteria---are constructible literals unless the current tool metadata explicitly requires a pre-existing reference.
11. For a call that authenticates, reads, changes, deletes, sends to, or relates a pre-existing record, use observation for every identity-bearing account value, credential, recipient/contact, identifier, handle, code, or path that must match that record. A caller typing the value does not make it independently constructible.
12. In an expected-success authentication or sign-in template, both the existing account identifier and its existing secret are observations; never invent either. Account creation instead constructs its new identifier and secret. Credential recovery observes the existing account and issued recovery code but constructs the replacement secret.
13. Do not use observation merely because another tool could output a schema-compatible value; the value must be tied to a particular pre-existing entity, credential, file, or state needed by this call.
14. Use omit only for an optional parameter that the template intentionally omits.
15. Omit is orthogonal to dependency: when an optional parameter is supplied and must reference a particular pre-existing value, use observation.
16. Across the proposed success templates, give every optional public parameter that a caller can construct at least one literal example; another success template may still omit it.
17. Literals are frozen proposals; ordinary propagation will not replace them.
18. Include at least one proposed success template.
19. Declare every reusable value with an explicit JSONPath export.
20. Caller-supplied arguments remain observable after the call; their mechanical $.parameter_name exports are normalized from the public signature.
21. Export every scalar leaf shown in the public success response; use [*] for array elements.
22. For every response or state_changes export, set propagate=true when its exact runtime value identifies, locates, describes, authenticates, or carries specific entity/content/state information that a later tool call could meaningfully reuse.
23. Every identifier or foreign key, opaque handle/token/code, address/contact, path/URL, timestamp/date, user-authored content, entity attribute, and monetary value must propagate, even when echoed or easy to replace with some other schema-valid value.
24. Set propagate=false only for generic acknowledgement/status/error text, boolean flags, aggregate counts/ratings/ranks, pagination/control metadata, or display-only summaries that neither identify nor describe a particular entity or content item.
25. Judge propagation from this tool alone and never from the runtime scalar's magnitude; a small integer or short string can still be an opaque resource reference.
26. Exports may select arguments, response, or state_changes and require a JSON Schema.
27. When same-tool feedback supports semantic names and exact paths for scalar state fields, export every such field; never return only a representative subset.
28. Same-tool JIT feedback is rendered exactly according to feedback_policy; use only the supplied representation to map meaningful field names to exact paths.
29. Every state export must name exactly one semantic field and use one explicit JSONPath.
30. A state_changes export must select reusable scalar leaves, not only a whole array or object container.
31. Never use recursive catch-all exports such as $..[*]; omit a hidden value when its field identity cannot be supported by the supplied context.
32. Never copy a concrete runtime scalar from invocation_result into a template, export name, path, or schema.
33. Do not export arbitrary descendants whose semantic identity is supported by neither public metadata nor supplied same-tool feedback.
34. Use standard JSON Schema Draft 2020-12 for observable success and failure outcomes.
35. Outcome schemas are over the normalized invocation record, never the raw response alone.
36. A success outcome schema must match top-level status success and put response constraints under output.
37. A failure outcome schema must match top-level status rejected or error and put response constraints under output.
38. The success and failure schemas must describe disjoint observable outcomes.
39. Use only this tool's public metadata and supplied same-tool feedback.
40. Do not infer another tool, graph, task, benchmark, fixture, or application rule.
\end{promptbox}
\captionof{figure}{Shared rules supplied through \texttt{contract\_rules} during \textsc{LIBRA} contract generation.}

\paragraph{Catalog-specific calling conventions.}
AppWorld uses the shared prompt unchanged.
SciToolEval and BFCL insert the following rules after Rule 1 and replace Rules 14, 15, 20, and 21 with their registry-specific versions.

\begin{promptbox}
# Inserted registry rules
- Return only fields defined by required_output_schema; metadata and calling-convention fields are input context, not contract output.
- {calling_convention_rule}
- Copy template argument keys exactly from public_tool_metadata.parameters[*].name; never invent an alias or use a parameter name found only in prose.
- If public_tool_metadata.parameters is empty, every template arguments object must be empty.
- Every literal must validate against that parameter's exact JSON Schema using native JSON types; numbers and booleans are not quoted strings.
- Never emit a long payload as a literal. For a sequence, document, file, blob, encoded value, or any input whose plausible example would be long, use observation instead; every literal must remain short enough to inspect on one line.
- Use the shortest literal that can plausibly succeed; when metadata states a minimum length, meet it without padding beyond that minimum.

# Registry replacements
- Use omit when a parameter is intentionally absent from this template; omit is a call choice, not a model-filled value.
- Omit is orthogonal to dependency: a field may be absent in one template and still require an upstream observation whenever supplied.
- Do not declare caller-argument exports; they are added mechanically from the public interface.
- For a structured public success response, do not repeat its scalar exports; exact public field labels and paths are added mechanically. For a scalar success response, declare exactly one response export at path \$ and name it with the shortest snake_case semantic value name explicitly supported by this tool's public function, description, or return documentation. Use response_value only when that metadata gives the scalar no semantic identity.

# Appended registry rules
- Keep outcome schemas minimal: constrain normalized status and output shape without copying the full public response object.
- Before returning, check every template key against public_tool_metadata.parameters[*].name, every required parameter against its template, and every literal against its listed schema.

# System-message suffix
Return compact JSON. Never write or repeat a long payload literal; represent that input as observation.
\end{promptbox}
\captionof{figure}{Registry-specific calling conventions used for SciToolEval and BFCL.}

For BFCL, the seventh inserted rule and the failure-outcome rule are replaced, and the final three rules below are appended.

\begin{promptbox}
- For a state-changing action, put first a success template that completes the operation described by the public metadata rather than a partial, no-op, or opposite transition. When the metadata describes a bounded degree, use its completing boundary; when a collection selects independently named components and lists the complete allowed set, include that full set. Prefer this one effect-complete template over extra reset or diversity templates.
- A failure outcome schema normally matches top-level status rejected or error. It may instead match top-level status success only when public metadata explicitly defines an output value as a business-level failure.
- A transport-success response can still report a business-level failure. Use an output const or enum as a success/failure discriminator only when this tool's public metadata explicitly defines those values.
- For a read, query, or status operation, a returned boolean is successful result data unless public metadata explicitly says one value means the operation failed; otherwise both true and false belong to the success outcome.
- For an action, authentication, or mutation whose public metadata explicitly defines success and failure flag values, constrain those values in disjoint success and failure schemas.
\end{promptbox}
\captionof{figure}{Additional contract-generation rules used for BFCL stateful tools.}

When feedback is provided, \texttt{feedback\_note} states that it is restricted to the current tool and that concrete runtime values may be used only to interpret field structure, never copied into persisted literals.
For the \texttt{complete-same-tool} policy used in all three benchmarks, the exact rule is:

\begin{promptbox}
Feedback is restricted to this tool. invocation_result is the complete return record for the call that triggered JIT repair; other feedback scalars remain structural. Use runtime values only to understand field structure and never copy a concrete feedback value into any persisted contract literal. Named state columns or keys are structural evidence: enumerate every scalar field whose semantic name and exact path the record supports.
\end{promptbox}
\captionof{figure}{Feedback instruction used when revising a contract after same-tool execution.}

The calling-convention placeholder is filled with exactly one of the following rules, according to the public interface:

\begin{promptbox}
This runtime receives one JSON object: template.arguments are exactly that object's public keys, with their original spelling.

This runtime uses the declared callable signature: template.arguments are exactly its public parameter names.
\end{promptbox}
\captionof{figure}{Calling-convention text selected according to the benchmark tool interface.}

\begin{promptbox}
# System
You enrich under-documented software tools for retrieval. Given a tool's name, signature, and docstring, output strict JSON with keys 'function_description' (1-2 sentences on what it does), 'when_to_use' (when a task needs this tool), 'limitations' (what it does not do / requires), and 'tags' (a list of short topical keywords). No prose.

# User
Tool:
{tool_name}{tool_signature}
Category: {tool_category}
{tool_docstring}

Return the JSON object.
\end{promptbox}
\captionof{figure}{Tool-REX initial documentation-generation prompt.}
\label{prompt:tool-rex-initial}

\begin{promptbox}
You will receive a copy of the API documentation.
Your tasks are strictly based on this API. A new "tool_profile" field
has been added to the documentation.
The goal is to make the overall documentation more accurate and
complete, but not inconsistent.
## Required Fields (Always included in tool_profile)
- "function": less than 20 words; clearly describe the tool's main
function using real terms.
- "tags": 3-5 lowercase keywords; no repetition or self-contradictions
; phrases covering the main topic, primary operations, and
synonyms.
## Optional Fields (Included only if explicitly supported by the
original documentation)
- "when_to_use": less than 20 words; actual usage scenarios.
- "example_usage": Maximum of 2 Include projects only if the actual
example can be constructed directly from the fields in the
original document. Each project may contain:
- "query": A natural or keyword-based question
- "api_call": An actual API usage call consistent with the tool's
actual parameters.
- "limitation": Included only if the original document explicitly
mentions limitations, rate limits, or special conditions.
## Important Rules
- Absolutely no fiction: Do not add functions, parameters, outputs,
limitations, or fields that do not exist in the original document.
- Consistency: If the original document is about finance, don't
mention weather. If it's about image processing, don't mention
stock data. Stick to the relevant fields.
- If optional fields are not explicitly supported, omit them entirely.
- Output only valid JSON objects, formatted as follows:
{
"tool_profile": {
"function": "...",
"tags": ["...", "..."],
"when_to_use": "...",
"example_usage": [
{
"query": "...",
"api_call": "..."
}],
"limitation": "..."
}}
Important:
- The output must be directly parsable via json.loads() .
- Do not include comments, instructions, or Markdown code fences.
- Do not use the <think> step.
- If the documentation does not explicitly support an optional field,
omit it rather than adding it yourself.
The "tool_profile" document is now generated based on the following
API:
{api_document}
\end{promptbox}
\captionof{figure}{Tool-REX profile-refinement prompt. Gemma-4-31B applies it to profiles rejected by the consistency judge.}
\label{prompt:tool-rex-refinement}

\begin{promptbox}
You will receive two inputs:
(1) The original API documentation.
(2) An expanded "tool_profile" derived from the documentation.
Your task is to judge whether the tool_profile
is consistent with the original documentation.
## Field Definitions
- "function": The main purpose of the tool.
- "tags": Keywords describing the domain and usage.
- "when_to_use": Scenarios where the tool should be applied.
- "example_usage": Example queries and corresponding API calls.
- "limitation": Constraints explicitly mentioned in the original
documentation.
## Important Rule
- Do not invent or assume information.
- Base your judgement only on the given API documentation.
Output only one word: "true" if the tool_profile is consistent,
otherwise "false".

API documentation:
{api_document}

tool_profile:
{tool_profile}
\end{promptbox}
\captionof{figure}{Tool-REX consistency-judgement prompt. It returns a binary decision before a generated profile is accepted.}
\label{prompt:tool-rex-judgement}

\begin{promptbox}
# System
You classify the data-flow dependency between two tools. Judge ONLY from their I/O schema (the signature gives REQUIRED INPUTS; the docstring 'Returns'/functionality gives the OUTPUT). A dependency exists when one tool's OUTPUT is exactly what the other tool needs as a REQUIRED INPUT (same type AND meaning). The direction points from the consumer to the producer: 'X depends on Y' means Y produces a value that X requires as input. Output only strict JSON with one key 'label'. No reasoning. Valid labels: a_depends_on_b, b_depends_on_a, no_dependency.

# User
Determine, from inputs/outputs only, whether one tool's output feeds a required input of the other. Ignore shared category, similar names, or merely related topics.

Worked example:
Tool A:
Tool ID: ConvertingPeptide2SMILES
Signature: (peptide: str)
Docstring: Convert a peptide sequence into its SMILES string. Returns: a SMILES string.
Tool B:
Tool ID: SMILESToInChI
Signature: (smiles: str)
Docstring: Convert a SMILES string into an InChI. Returns: an InChI string.
Analysis: A outputs a SMILES string; B's required input is a SMILES string. So B consumes A's output -> B depends on A.
Answer: {"label":"b_depends_on_a"}

If neither tool's output is a required input of the other, the answer is {"label":"no_dependency"}.

Now classify this pair:
Tool A:
Tool ID: {tool_a_id}
Signature: {tool_a_signature}
Docstring:
{tool_a_docstring}

Tool B:
Tool ID: {tool_b_id}
Signature: {tool_b_signature}
Docstring:
{tool_b_docstring}

Return exactly one JSON object such as {"label":"no_dependency"}.
\end{promptbox}
\captionof{figure}{TGR dependency judgement prompt. It assigns a directed dependency label or \texttt{no\_dependency} to every unordered tool pair.}
\label{prompt:tgr-dependency}

\begin{promptbox}
# System
Extract a compact knowledge graph from one tool description. Return only JSON with a 'relationships' array. Every item must contain the string fields 'head', 'relationship', and 'tail'. Use concise, consistent entity names; split compound claims; do not duplicate triples or create self-relations.

# User
The exact tool title is {tool_name}. Extract head-relation-tail triples from the metadata below. Use only these relations when applicable: has_line_of_business, has_entity, has_parameter, contains, related_to, used_for, has_attribute, associated_with, managed_by, part_of, required_for, depends_on, produces, receives_from, involved_in, reports_to, responsible_for, affects, includes. Link the title to its category with has_line_of_business, to each input parameter with has_parameter, and to other salient concepts with has_entity. The exact tool title may occur only in those three relations. Preserve the exact title whenever it is a head.

Scenario title: {tool_name}
Line of business: {category}
Function: {function}
Description: {description}
Input parameters: {parameter_names}

Output schema: {"relationships":[{"head":"...","relationship":"...","tail":"..."}]}
\end{promptbox}
\captionof{figure}{Ego-Trip graph-extraction prompt adapted to each benchmark catalog.}
\label{prompt:egotrip-extraction}

All agentic retrieval experiments use the \texttt{plain\_fulltask} condition below.

\begin{promptbox}
# System
You are designed to help with a variety of tasks, from answering questions to providing summaries to other types of analyses.

## Tools
You have access to a wide variety of tools. You are responsible for using the tools in any sequence you deem appropriate to complete the task at hand. This may require breaking the task into sub-tasks and using different tools to complete each sub-task.

You have access to the following tools:
{tool_desc}

## Output Format
Please answer in the same language as the question and use the following format:

Thought: The current language of the user is: (user's language). I need to use a tool to help me answer the question.
Action: tool name (one of {tool_names}) if using a tool.
Action Input: the input to the tool, in a JSON format representing the kwargs (e.g. {"input": "hello world", "num_beams": 5})

Please ALWAYS start with a Thought. NEVER surround your response with markdown code markers. You may use code markers within your response if you need to. Please use a valid JSON format for the Action Input. Do NOT do this {'input': 'hello world', 'num_beams': 5}.

If this format is used, the tool will respond in the following format:

Observation: tool response

You should keep repeating the above format till you have enough information to answer the question without using any more tools. At that point, you MUST respond in one of the following two formats:

Thought: I can answer without using any more tools. I'll use the user's language to answer
Answer: [your answer here, in the same language as the user's question]

Thought: I cannot answer the question with the provided tools.
Answer: [your answer here, in the same language as the user's question]

## Current Conversation
Below is the current conversation consisting of interleaving human and assistant messages.

# User
Given the following complex task: "{task}" and the list of tools in the context, select the best tools to solve the complex task. You must obtain that context by calling "tool_retriever" with a suitable query before selecting tools.
\end{promptbox}
\captionof{figure}{ToolReAGt agent prompt used in all agentic retrieval experiments.}
\label{prompt:toolreagt}

The retriever is exposed as \texttt{tool\_retriever} with one string argument, \texttt{input}.
The \texttt{tool\_desc} placeholder uses the corresponding description below.
At later turns, LlamaIndex appends the preceding \texttt{Thought}, \texttt{Action}, and JSON \texttt{Observation} records to the same conversation.

\begin{promptbox}
# QTS, Tool-REX, and TGR
Search the complete tool registry with a focused natural-language query and return the top ranked tool descriptions.

# Ego-Trip
Search the tool knowledge graph with a focused natural-language query and return tools from the matched ego graphs.

# LIBRA
Search an execution-validated dependency library with a focused natural-language query and return a ranked tool list.
\end{promptbox}
\captionof{figure}{Retriever descriptions supplied to ToolReAGt for the compared retrieval backends.}

\section{Execution Validation Ablation}
\label{app:ablation}

\begin{table}[htbp!]
\centering
\caption{Ablation of graph quality with fixed dense retrieval.
All variants use the same encoder over original tool metadata, differing
only in the retrieval graph.}
\label{tab:ablation}
\resizebox{\columnwidth}{!}{
\begin{tabular}{lccc}
\toprule
 & QTS & Schema-Only & \textsc{LIBRA} \\
\midrule
AppWorld@15 R / NDCG & 28.7 / 32.8 & 29.2 / 33.1 & \textbf{40.7} / \textbf{39.1} \\
SciToolEval@10 R / NDCG & 84.1 / 83.9 & 84.1 / 83.9 & \textbf{91.1} / \textbf{86.6} \\
BFCL@15 R / NDCG & 81.8 / 76.2 & 81.4 / 76.0 & \textbf{86.4} / \textbf{78.4} \\
\bottomrule
\end{tabular}}
\end{table}

We examine the effect of execution-based filtering by fixing the dense retriever and varying only the retrieval graph used for score propagation.

All three variants use Qwen-3 Embedding-0.6B over the original tool metadata, as in QTS, and no reranking is applied.
The first variant (QTS) performs no propagation. The second propagates scores over all schema-compatible edges, i.e., the candidate graph before execution filtering. 
LIBRA propagates only over edges retained through successful execution in the target runtime.

Table~\ref{tab:ablation} shows that schema-compatible edges alone are insufficient. 
Propagation over the execution-filtered graph consistently outperforms both QTS and propagation over the unfiltered schema-compatible graph.

\section{Library Construction Statistics}
\label{sec:construction-statistics}

Table~\ref{tab:construction-statistics} reports the scale and runtime cost of \textsc{LIBRA}'s ahead-of-time library construction.
For each tool, DeepSeek-V4-Flash generates an initial executable contract with temperature zero, and \textsc{LIBRA} permits at most two feedback based revisions.
For AppWorld, SciToolEval, and BFCL, respectively, 218, 48, and 37 contracts required at least one revision.

The resulting contracts define candidate compositions between schema compatible provider exports and consumer inputs.
A single provider consumer tool pair can contribute multiple candidate compositions when different exports or inputs are involved.
Across the three benchmarks, \textsc{LIBRA} generates between 12,762 and 581,690 candidate compositions before runtime filtering and retains between 672 and 1,946 execution supported relations.

Each library construction job was configured with 96 CPU workers.
The AppWorld job requested 384~GB of memory, whereas the SciToolEval and BFCL jobs each requested 128~GB.
Each job additionally used one L40S GPU to serve field embeddings.
The recorded wall clock times were 260 minutes 40 seconds for AppWorld, 20 minutes 45 seconds for SciToolEval, and 4 minutes 28 seconds for BFCL.

The total construction time depends on two distinct factors.
The first is the number of candidate compositions that must be tested.
This quantity depends not only on catalog size, but also on the numbers of exported provider values, consumer inputs, and schema compatible matches between them.
For example, AppWorld and SciToolEval contain similar numbers of tools, 473 and 477, but produce 581,690 and 163,608 candidate compositions, respectively.

The second factor is the cost of validating each candidate composition.
A candidate composition is not equivalent to a single sandbox execution: its validation can invoke the provider, the consumer, and one or more control calls.
The invoked tools can also differ in their internal execution cost, including both constant overhead and asymptotic complexity with respect to their own input sizes.
In addition, the three benchmarks use different executable backends.
Consequently, candidate count alone does not determine wall clock time.
When normalized by candidate count, the recorded times correspond to approximately 26.9, 7.6, and 21.0 seconds per 1,000 candidate compositions for AppWorld, SciToolEval, and BFCL, respectively.
AppWorld takes approximately 12.6 times as long as SciToolEval despite generating approximately 3.6 times as many candidate compositions.

In the worst case, the number of candidate compositions is quadratic in the catalog size, with additional multiplicative factors from the numbers of provider exports and consumer inputs.
Schema compatibility restricts the number of compositions considered in practice.

None of the three benchmarks provides complete provider consumer dependency annotations. We therefore do not construct surrogate edge level ground truth or claim precision against a canonical dependency graph.
Here, \emph{runtime validated} means that a retained relation passed
the corresponding composition test in the benchmark's executable
environment: AppWorld's local stateful runtime, SciToolEval's service
based tool runtime, or BFCL's released executable backend.
Runtime validation provides execution evidence under the tested
contract and environment state rather than a universal guarantee over
all possible inputs and states.

\begin{table}[t]
\centering
\caption{Library construction statistics.
Candidate compositions are schema compatible combinations of provider
exports and consumer inputs before runtime filtering.
A single tool pair may contribute multiple candidate compositions.
Relation counts report value based relations and state dependencies
included in the benchmark specific retrieval graph.
Normalized time is calculated from the recorded wall clock time and
candidate composition count.}
\label{tab:construction-statistics}
\resizebox{\columnwidth}{!}{
\begin{tabular}{lrrr}
\toprule
 & AppWorld & SciToolEval & BFCL \\
\midrule
Catalog tools
& 473 & 477 & 128 \\
Candidate compositions
& 581{,}690 & 163{,}608 & 12{,}762 \\
Retained relations (value based / state)
& 1{,}946 / 0 & 732 / 0 & 595 / 77 \\
Wall clock (min:s)
& 260:40 & 20:45 & 4:28 \\
Seconds per 1,000 candidates
& 26.9 & 7.6 & 21.0 \\
\bottomrule
\end{tabular}}
\end{table}

\section{Dataset Processing and Baseline Adaptation}
\label{sec:data}

\paragraph{Benchmark conversion and curation.}
We convert each benchmark into a common retrieval format containing
a query, a shared tool catalog, and a gold required-tool annotation.
When the released annotation contains an ordered call sequence, we
preserve it as \texttt{tool\_path}; retrieval evaluation is defined
over the set of distinct tools in that sequence.

\textbf{For AppWorld,} we evaluate on the 90 training and 57 development tasks with public required-API annotations; hidden test annotations are unavailable.

\textbf{For SciToolEval,} the released files contain 530 tasks: 150 level-1 and 380 level-2. We evaluate a frozen subset of 270 tasks and exclude the remaining 260. Each retained task keeps its released split name and row index, and we verify that its question and ordered gold tool path match the corresponding released record exactly. The same subset is used for all retrieval methods.

Tasks are excluded when they involve tools that cannot be executed in the runtime. Inspecting the released implementation, we identify one important source of incompatibility: the \texttt{IsValidInChIKey}, \texttt{InChIKeyToSMILES}, \texttt{InChIKeyToInChI}, and \texttt{InChIKeyToMOL} wrappers all invoke the legacy ChemSpider InChI web service. On such non-executable tools LIBRA induces no composition relations and reduces to the dense baseline, so the exclusion confers no advantage specific to LIBRA over the non-execution baselines.

\textbf{For BFCL,} we use the 200 tasks from the BFCL v4 \texttt{multi\_turn\_base}
split and the 128 functions in its released multi-turn function
registry. Each task is matched by identifier to one record in
\texttt{possible\_answer/BFCL\_v4\_multi\_turn\_base.json}. We read
the record's \texttt{ground\_truth} field as an ordered list of turns,
where each turn contains the official function calls for that turn.
We flatten these calls in chronological order and map every raw
function name to its unique fully qualified API name.

The stored \texttt{tool\_path} preserves both call order and repeated
calls. Retrieval relevance, however, is defined over the set of
distinct fully qualified functions in this sequence; repeated
invocations of the same function therefore receive no additional
relevance credit. We do not union or intersect alternative workflows,
nor do we select among multiple candidate answers: the converter
consumes the single \texttt{ground\_truth} object associated with each task.

The retrieval query is formed by collecting every non-empty user
message from all turns, preserving the original turn order, and
joining the messages with newline separators. Assistant messages,
tool outputs, function arguments, and the released ground-truth calls are not included in the retrieval query.

\paragraph{Retrieval metrics.}
Let \(G_i\) denote the set of distinct gold tools for query \(i\),
and let
\[
R_i^K = (r_{i1}, \ldots, r_{iK})
\]
denote the ranked top-\(K\) retrieval result. We compute
\[
\operatorname{Precision@K}_i =
\frac{1}{K}
\sum_{j=1}^{K}
\mathbb{I}[r_{ij} \in G_i],
\]
and
\[
\operatorname{Recall@K}_i =
\frac{1}{|G_i|}
\sum_{j=1}^{K}
\mathbb{I}[r_{ij} \in G_i].
\]

For NDCG, tool relevance is binary:
\[
\operatorname{rel}_{ij} =
\mathbb{I}[r_{ij} \in G_i].
\]
We compute
\[
\operatorname{DCG@K}_i =
\sum_{j=1}^{K}
\frac{\operatorname{rel}_{ij}}{\log_2(j+1)}
\]
and
\[
\operatorname{IDCG@K}_i =
\sum_{j=1}^{\min(K, |G_i|)}
\frac{1}{\log_2(j+1)}.
\]
Then
\[
\operatorname{NDCG@K}_i =
\frac{\operatorname{DCG@K}_i}
     {\operatorname{IDCG@K}_i}.
\]

All reported Precision, Recall, and NDCG values are unweighted averages; therefore, tasks with larger gold sets do not receive greater weight. Repeated calls to the same gold function, including repeated calls in BFCL workflows, are collapsed into one relevant tool before metric computation.

LIBRA is training-free and does not access task instructions, gold
tool sets, or split labels during library construction.
Required-tool annotations are used only after retrieval for evaluation. We perform no hyperparameter selection using task-level annotations from AppWorld, SciToolEval, or BFCL.

All embedding-based methods use the same
Qwen-3 Embedding-0.6B encoder. For the standard dense representation,
we embed the original serialized tool metadata. For Tool-REX,
DeepSeek-V4-Flash generates a structured profile for each tool, and
Gemma-4-31B performs one consistency check against the source metadata
before the profile is embedded using the same encoder.

The original TGR implementation and trained dependency discriminator are not publicly available; we therefore implement a TGR-style adaptation in which DeepSeek-V4-Flash labels dependencies for tool pairs from their metadata, after which we apply TGR's parameter-free graph propagation to the frozen tool embeddings. For Ego-Trip, we adapt its metadata based prompting procedure to each benchmark, use the dense retrieval tool in our experiments, and keep the remaining prompts the same. All prompts are available in Appendix~\ref{sec:prompts}.

\section{Retrieval Sensitivity}
\label{sec:sensitivity}

We vary three retrieval-time settings while holding the executable
library, embedding model, and evaluation cutoff fixed (Table~\ref{tab:sensitivity}).

\begin{table}[htbp!]
\centering
\small
\caption{Sensitivity to retrieval-time settings.
The induced library is fixed throughout. Values are R / NDCG (\%).}
\label{tab:sensitivity}
\resizebox{\columnwidth}{!}{
\begin{tabular}{llccc}
\toprule
Setting & Variant & AppWorld@15 & SciToolEval@10 & BFCL@15 \\
\midrule
\multirow{4}{*}{\shortstack[l]{Provider\\selection}}
& Field-best only   & 40.3 / 38.9 & \textbf{92.0} / \textbf{87.0} & 86.0 / 78.2 \\
& Dense-best only    & 39.7 / 38.5 & 88.0 / 85.3 & 84.8 / 77.7 \\
& Both (default)     & \textbf{40.7} / \textbf{39.1} & 91.1 / 86.6 & \textbf{86.4} / \textbf{78.4} \\
& All retained      & 38.6 / 38.5 & 76.4 / 75.0 & 85.3 / 78.0 \\
\midrule
\multirow{3}{*}{Seed $K_s$}
& 10                 & 40.8 / \textbf{39.6} & \textbf{91.9} / \textbf{86.9} & 86.2 / 78.5 \\
& 15 (default)       & \textbf{40.7} / 39.1 & 91.1 / 86.6 & \textbf{86.4} / \textbf{78.4} \\
& 20                 & 40.6 / 39.0 & 91.3 / 86.7 & 86.2 / 78.3 \\
\midrule
\multirow{2}{*}{Hop $h$}
& 1 (default)        & \textbf{40.7} / \textbf{39.1} & \textbf{91.1} / \textbf{86.6} & \textbf{86.4} / 78.4 \\
& 2                  & 40.4 / 39.1 & 88.8 / 85.5 & \textbf{86.4} / \textbf{78.5} \\
\bottomrule
\end{tabular}}
\end{table}

\paragraph{Provider selection.}
Retaining all providers connected by the retained relations degrades
Recall sharply on SciToolEval (76.4 vs.\ 91.1), showing that unpruned
retained relations can introduce retrieval noise.
Field-best and dense-best selection each recover most of the gain
independently; combining them (default) yields the most consistent
results across benchmarks.

\paragraph{Seed size.}
Performance is stable across $K_s \in \{10, 15, 20\}$, with
fluctuations below one percentage point on all benchmarks.

\paragraph{Hop depth.}
Extending propagation to two hops slightly reduces Recall on
SciToolEval ($-2.3$) while leaving AppWorld and BFCL essentially
unchanged.

\section{Implementation Details}
BM25 was implemented directly rather than through an external retrieval
package. We lowercase text and tokenize it using the regular expression
\texttt{[A-Za-z0-9\_]+}, with $k_1=1.5$ and $b=0.75$.
Precision, Recall, and NDCG were also implemented directly according to
the definitions in Appendix~D.

Dense rankings use the released \texttt{Qwen/Qwen3-Embedding-0.6B} representations. 
Tool and query embeddings are $\ell_2$-normalized, cosine similarity is computed by NumPy matrix multiplication, and rankings are obtained with
\texttt{argsort}; no nearest-neighbor package is used~\citep{numpy}.

For Figure~\ref{fig:dependency-semantic-gap}, we project each benchmark's tool embeddings independently using \texttt{sklearn.manifold.TSNE} with two components, cosine distance, perplexity $30$, automatic learning rate, random initialization, random seed $17$, and $1,500$ iterations~\citep{tsne,scikit-learn}. 
Model inference uses vLLM~\citep{vllm}.
\onecolumn
\newpage
\section{Agentic Retrieval across Iterations}
\label{sec:agentic-iterations}

\begin{figure}[htbp!]
    \centering
    \begin{subfigure}[t]{\columnwidth}
        \centering
        \includegraphics[
            width=\linewidth
        ]{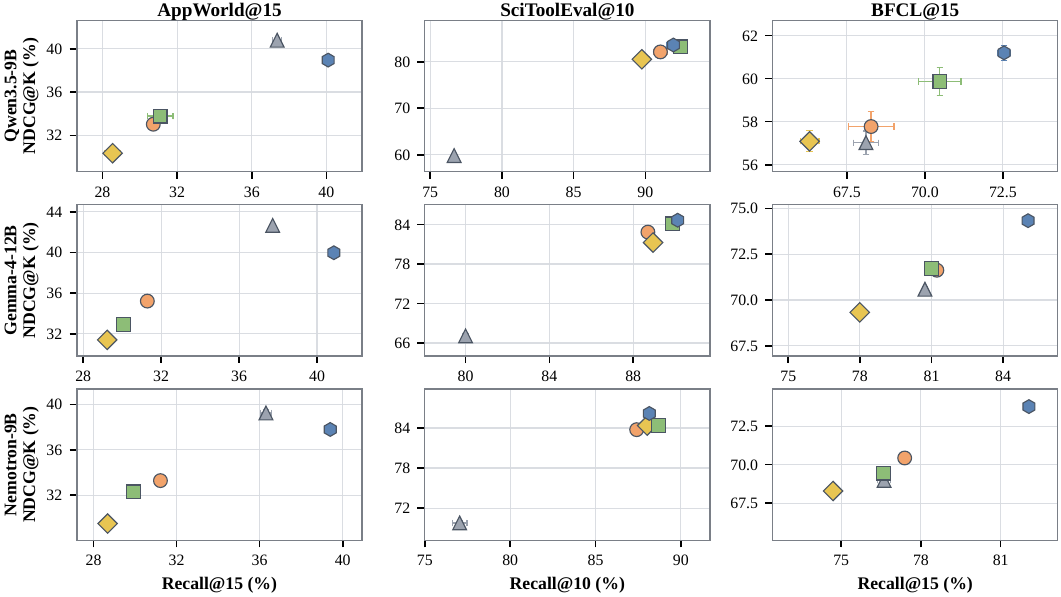}
        \caption{After 1 iteration.}
        \label{fig:agentic-iter1}
    \end{subfigure}
    \begin{subfigure}[t]{\columnwidth}
        \centering
        \includegraphics[
            width=\linewidth
        ]{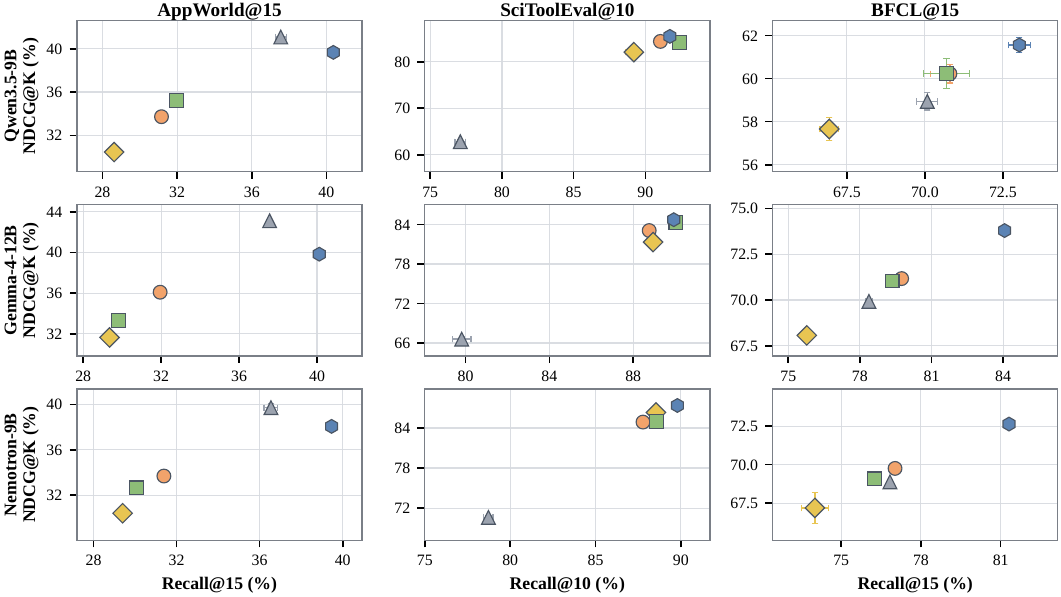}
        \caption{After 3 iterations.}
        \label{fig:agentic-iter3}
    \end{subfigure}

    \includegraphics[
        width=0.58\columnwidth
    ]{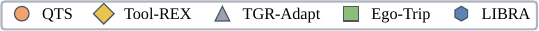}

    \caption{
        Agentic retrieval across ToolReAGt iterations.
        Each subfigure reports Recall and NDCG for three agent LLMs on
        AppWorld, SciToolEval, and BFCL at their benchmark-specific
        evaluation depths.
        Results show the mean over three random seeds, and error bars show
        standard error.
    }
    \label{fig:agentic-iteration-curves}
\end{figure}

\begin{figure}[htbp!]
    \ContinuedFloat
    \centering

    \setcounter{subfigure}{2}

    \begin{subfigure}[t]{\columnwidth}
        \centering
        \includegraphics[
            width=\linewidth
        ]{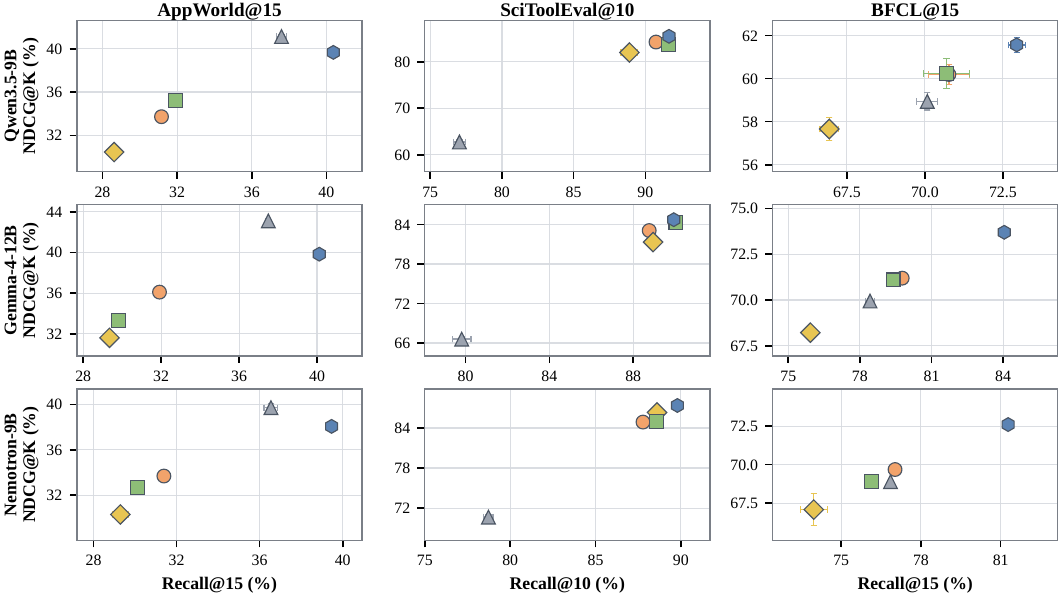}
        \caption{After 5 iterations.}
        \label{fig:agentic-iter5}
    \end{subfigure}

    \par

    \begin{subfigure}[t]{\columnwidth}
        \centering
        \includegraphics[
            width=\linewidth
        ]{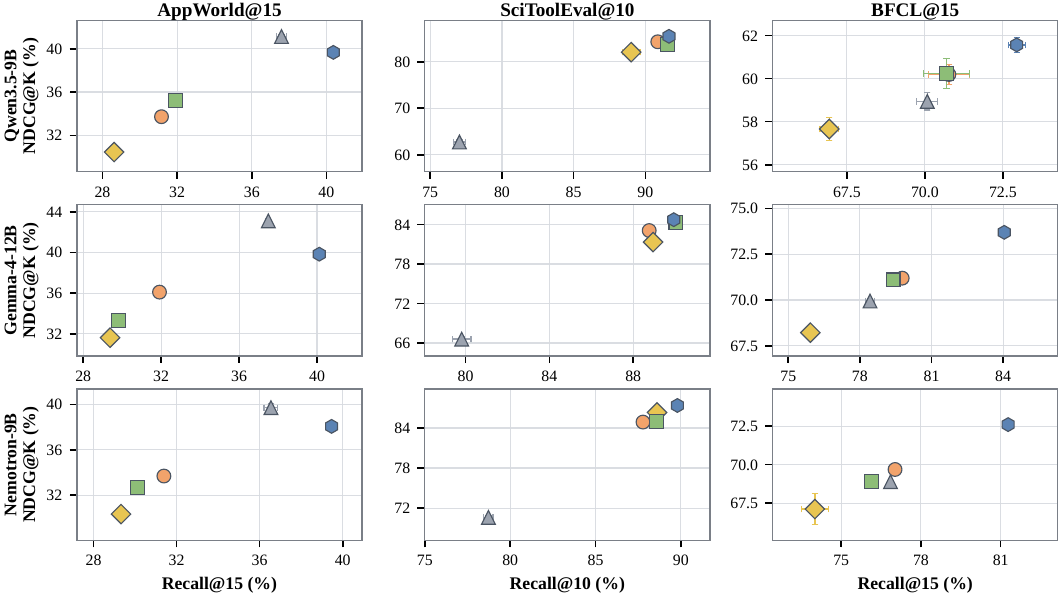}
        \caption{After 7 iterations.}
        \label{fig:agentic-iter7}
    \end{subfigure}

    \par

    \includegraphics[
        width=0.58\columnwidth
    ]{figures/fig_agentic_retrieval_legend.pdf}
    \caption[]{
        Agentic retrieval across ToolReAGt iterations
        \textnormal{(continued).}
    }
\end{figure}

Figure~\ref{fig:agentic-iteration-curves} shows additional results for agentic retrieval across different iterations. 
Each subfigure reports Recall and NDCG for three agent LLMs on AppWorld, SciToolEval, and BFCL at their benchmark-specific evaluation depths. 
Results show the mean over three random seeds, and error bars show standard error. LIBRA achieves the highest Recall in 30 settings and the highest NDCG in 24 across all 36 settings. Relative to standard dense retriever QTS, LIBRA improves Recall and NDCG by 4.5 and 3.0 points on average.

\section{Algorithms}
\label{sec:algorithms}

Algorithms~\ref{alg:library-construction}--\ref{alg:graph-propagation}
summarize the key steps in library construction and graph propagation.
Here, \(m_i\) is tool \(t_i\)'s public
metadata, and \(\operatorname{Run}_{\sigma}\) starts from the state \(\sigma\).
The contract is used by \(\operatorname{Extract}\) and
\(\operatorname{succ}/\operatorname{fail}\).
Let \(f_{\mathrm{out}}\) and \(f_{\mathrm{in}}\) denote output- and
input-field names.  A work item \((\sigma,t_i,\pi_i,\beta)\) runs candidate
call \(\pi_i\in\Pi_i\) of tool \(t_i\) from state \(\sigma\).
The binding map \(\beta\) is empty for an initial call; otherwise,
\(\beta(f_{\mathrm{in}})=(t_p,f_{\mathrm{out}},v)\) fills
\(f_{\mathrm{in}}\) with value \(v\) from provider \(t_p\)'s output field
\(f_{\mathrm{out}}\).
Field alignment passes when
\(\cos(h_f(f_{\mathrm{out}}),h_f(f_{\mathrm{in}}))>\tau\), where \(h_f\)
is the dense encoder for field names.
We write \(\mathcal E_{\mathrm{prov}}\), \(\mathcal E_{\mathrm{comp}}\),
and \(\mathcal E_{\mathrm{state}}\) for input-provenance relations,
compatibility relations, and state dependencies, respectively.

\SetAlgoNlRelativeSize{-1}
\begin{algorithm2e}[H]
\small
\DontPrintSemicolon
\KwIn{Tool catalog \(\mathcal{T}\), public metadata \(\{m_i\}_{i=1}^{n}\), runtime \(\mathcal{R}\), field encoder \(h_f\), and threshold \(\tau\)}
\KwOut{Executable library \(\mathcal{L}=(\mathcal{T},\mathcal{C},\mathcal{E})\)}
\(\mathcal{C},\mathcal{W}\leftarrow\varnothing,\varnothing\)\;
\(\mathcal E_{\mathrm{prov}},\mathcal E_{\mathrm{comp}},
\mathcal E_{\mathrm{state}}\leftarrow\varnothing,\varnothing,\varnothing\)\;
\ForEach{\(t_i \in \mathcal{T}\)}{
    \(c_i=(\Pi_i,\mathcal{X}_i,S_i^{+},S_i^{-})
    \leftarrow \operatorname{LLMGenerate}(m_i)\)\;
    \(b_i \leftarrow 2,\quad
    \mathcal{C} \leftarrow \mathcal{C}\cup\{c_i\}\)\;
}
\(\sigma_0\leftarrow\operatorname{Fresh}(\mathcal{R})\)\;
\ForEach{\(t_i\in\mathcal T,\ \pi\in\Pi_i\)}{
    Add \((\sigma_0,t_i,\pi,\varnothing)\) to \(\mathcal W\)\;
}
\While{\(\mathcal{W}\neq\varnothing\)}{
    \((\sigma,t_i,\pi_i,\beta)\leftarrow\operatorname{Pop}(\mathcal{W})\)\;
    \(\widehat\pi_i\leftarrow\operatorname{Fill}(\pi_i,\beta)\)\;
    \((r_{\mathrm{use}},\sigma')\leftarrow
    \operatorname{Run}_{\sigma}(t_i,\widehat\pi_i)\)\;

    \While{\(\neg[\operatorname{succ}(r_{\mathrm{use}})\lor
    \operatorname{fail}(r_{\mathrm{use}})]\land b_i>0\)}{
        \(c_i\leftarrow\operatorname{LLMReviseContract}
        (t_i,m_i,r_{\mathrm{use}})\)\;
        \(b_i\leftarrow b_i-1\)\;
    }

    \If{\(\operatorname{succ}(r_{\mathrm{use}})\)}{
        \ForEach{\(\beta(f_{\mathrm{in}})=(t_p,f_{\mathrm{out}},v)\)}{
            \(\tilde v\leftarrow
            \text{a test value not produced by }t_p\)\;
            \((r_{\mathrm{alt}},\_)\leftarrow
            \operatorname{Run}_{\sigma}(t_i,
            \widehat\pi_i[f_{\mathrm{in}}\leftarrow\tilde v])\)\;
            \uIf{\(\operatorname{fail}(r_{\mathrm{alt}})\)}{
                \(\mathcal E_{\mathrm{prov}}\leftarrow
                \mathcal E_{\mathrm{prov}}\cup
                \{(t_p,f_{\mathrm{out}},t_i,f_{\mathrm{in}})\}\)\;
            }
            \ElseIf{\(\operatorname{succ}(r_{\mathrm{alt}})\) and
            \(\cos(h_f(f_{\mathrm{out}}),h_f(f_{\mathrm{in}}))>\tau\)}{
                \(\mathcal E_{\mathrm{comp}}\leftarrow
                \mathcal E_{\mathrm{comp}}\cup
                \{(t_p,f_{\mathrm{out}},t_i,f_{\mathrm{in}})\}\)\;
            }
        }

        \ForEach{\((f_{\mathrm{out}},v)\in
        \operatorname{Extract}(\mathcal X_i,\widehat\pi_i,r_{\mathrm{use}})\)}{
            Add candidate calls using \(v\) as an input to \(\mathcal W\)\;
        }
        \If{\(\sigma'\neq\sigma\)}{
            \ForEach{\(t_c\in\mathcal T,\ \pi_c\in\Pi_c\)}{
                \((r_{\mathrm{before}},\_)\leftarrow
                \operatorname{Run}_{\sigma}(t_c,\pi_c)\)\;
                \((r_{\mathrm{after}},\_)\leftarrow
                \operatorname{Run}_{\sigma'}(t_c,\pi_c)\)\;
                \If{\(\operatorname{fail}(r_{\mathrm{before}})\) and
                \(\operatorname{succ}(r_{\mathrm{after}})\)}{
                    \(\mathcal E_{\mathrm{state}}\leftarrow
                    \mathcal E_{\mathrm{state}}\cup\{(t_i,t_c)\}\)\;
                }
                Add \((\sigma',t_c,\pi_c,\varnothing)\) to \(\mathcal W\)\;
            }
        }
    }
}
\(\mathcal E\leftarrow\mathcal E_{\mathrm{prov}}\cup
\mathcal E_{\mathrm{comp}}\cup\mathcal E_{\mathrm{state}}\)\;
\Return \((\mathcal{T},\mathcal{C},\mathcal{E})\)\;
\caption{Library construction.}
\label{alg:library-construction}
\end{algorithm2e}

\begin{algorithm2e}[H]
\small
\DontPrintSemicolon
\KwIn{Query \(q\), library \(\mathcal{L}\), encoders \(h,h_f\), seed size \(K_s\), and candidate size \(K_c\)}
\KwOut{Retrieved candidate set \(\mathcal{T}_q\)}
\ForEach{\(t\in\mathcal{T}\)}{
    \(s_q(t) \leftarrow [1+\cos(h(q),h(t))]/2\)\;
    \(s'_q(t) \leftarrow s_q(t)\)\;
}
\(S_q \leftarrow \operatorname{TopK}(\mathcal{T},s_q,K_s)\)\;
\ForEach{seeded consumer \(t\in S_q\)}{
    \(\widehat P(t)\leftarrow
    \{u:(u,t)\in\mathcal E_{\mathrm{state}}\}\)\;
    \ForEach{input-field name \(f_{\mathrm{in}}\) of \(t\)}{
        \(F(t,f_{\mathrm{in}}) \leftarrow
        \{(u,f_{\mathrm{out}}):(u,f_{\mathrm{out}},t,f_{\mathrm{in}})
        \in\mathcal E_{\mathrm{prov}}\cup\mathcal E_{\mathrm{comp}}\}\)\;
        \(P(t,f_{\mathrm{in}})\leftarrow
        \{u:\exists f_{\mathrm{out}},\ (u,f_{\mathrm{out}})\in
        F(t,f_{\mathrm{in}})\}\)\;
        \If{\(P(t,f_{\mathrm{in}})\neq\varnothing\)}{
            \((u_{\mathrm{field}},f_{\mathrm{out}}^*) \leftarrow
            \arg\max_{(u,f_{\mathrm{out}})\in F(t,f_{\mathrm{in}})}
            \cos(h_f(f_{\mathrm{out}}),h_f(f_{\mathrm{in}}))\)\;
            \(u_{\mathrm{dense}} \leftarrow
            \arg\max_{u\in P(t,f_{\mathrm{in}})}s_q(u)\)\;
            \(\widehat P(t)\leftarrow\widehat P(t)\cup
            \{u_{\mathrm{field}},u_{\mathrm{dense}}\}\)\;
        }
    }
}
\ForEach{provider \(u\in\mathcal{T}\)}{
    \(C_q(u) \leftarrow
    \{t\in S_q:u\in\widehat P(t)\}\)\;
    \If{\(C_q(u)\neq\varnothing\)}{
        \(g_q(u) \leftarrow
        \bigl(s_q(u)+\sum_{t\in C_q(u)}s_q(t)\bigr)/
        \bigl(1+|C_q(u)|\bigr)\)\;
        \(s'_q(u) \leftarrow \max\{s_q(u),g_q(u)\}\)\;
    }
}
\(\mathcal{T}_q \leftarrow \operatorname{TopK}(\mathcal{T},s'_q,K_c)\)\;
\Return \(\mathcal{T}_q\)\;
\caption{One-hop retrieval with graph propagation.}
\label{alg:graph-propagation}
\end{algorithm2e}

\section{Acknowledgments}
Zekai Li and Zhenke Liu contributed equally so they can prioritize their name in their resume.


\end{document}